# Nonequilibrium Quasiparticle Dynamics in a MoRe-Based Superconducting Resonator under Infrared Excitation

**O. A. Kalenyuk[1,2 a)], S. I. Futimsky[1,2], I. A. Martynenko[1,2], A. P. Shapovalov[1,2], O. O. Boliasova[1,2], V. I. Shnyrkov[1,2], A. L. Kasatkin[1,2], A. A. Kordyuk[1,2]**

**AFFILIATIONS**

[1]G. V. Kurdyumov Institute for Metal Physics, N.A.S. of Ukraine, Kyiv, 03142, Ukraine

[2]Kyiv Academic University, Kyiv, 03142, Ukraine

[a)]**Author to whom correspondence should be addressed:** oleksii.kaleniuk@gmail.com

**ABSTRACT**

The response of a MoRe-based microstrip superconducting resonator operating near 5 K to pulsed infrared irradiation is investigated, and the underlying physical mechanisms are analyzed. The device exhibits a pronounced nonlinear response dominated by nonequilibrium quasiparticle dynamics rather than uniform thermal heating. Infrared pulses produce strong distortions of the resonance curve and a transient decrease in the resonance frequency, consistent with increased kinetic inductance caused by quasiparticle generation. The frequency shift scales approximately linearly with absorbed power, whereas the dissipation response saturates at higher powers, indicating the formation of a nonequilibrium steady-state quasiparticle population. These observations demonstrate a transition from a linear pair-breaking regime to a saturated dissipation regime, likely associated with a quasiparticle relaxation bottleneck. The results highlight the relevance of nonequilibrium processes in MoRe and confirm its potential for microwave kinetic-inductance detector applications.

## I. INTRODUCTION

A wide variety of highly sensitive infrared (IR) detectors have been developed based on thin superconducting films. Two principal detection mechanisms are predominantly employed. The first relies on the bolometric effect, i.e., a sharp increase in the resistance of a narrow superconducting nanowire caused by the formation of a photon-induced hotspot. Such detectors are capable of resolving individual photons[1]. Their operation requires biasing the nanowire at a current close to the critical current and monitoring the onset of a finite voltage. This approach, however, necessitates a large number of electrical connections, particularly in the case of detector arrays. Additional drawbacks include the inability to detect radiation while the nanowire is in the normal state and the existence of a dead-time interval required for the system to recover its initial superconducting state. Thus, despite their excellent sensitivity, such detectors typically exhibit limited operational speed. The second approach is based on monitoring changes in the kinetic inductance of a superconducting film induced by photon absorption[2,3]. In this case, large arrays of superconducting resonators with distinct resonance frequencies are coupled to a single microwave feedline, significantly simplifying system architecture and enabling real-time two-dimensional imaging of incident electromagnetic radiation. The minimum detectable photon energy $h\nu=2\Delta$ is determined by the superconducting energy gap $\Delta$, which in low-temperature superconductors is typically on the order of several meV, and can be even lower in certain materials. For example, aluminum exhibits an energy gap of approximately $\Delta_{Al}$=0.18 meV. Microwave kinetic-inductance detectors (MKIDs) have found extensive application in astrophysics, in particular for sky mapping at millimeter and sub-millimeter wavelengths[4]. Their operation, however, requires deep-cryogenic temperatures.

For a wide range of emerging applications—such as medical diagnostics, security systems, and detection of radiation in the terahertz and IR spectral ranges under high background illumination—the dynamic range and response speed are often more critical than ultimate sensitivity. In such cases, MKIDs operating at temperatures of a few kelvin may provide significant advantages. For example, Ref. 5 demonstrated the operating principles of a submillimeter-wave detector based on a superconducting kinetic-inductance thermometer at temperatures around 6 K.

Usually, MKIDs rely on thin films of low-temperature superconductors that provide high-quality resonators with a strong kinetic inductance contribution to the electrodynamic response. The most widely used materials include titanium nitride (TiN),[6] Ti/TiN multilayers, niobium nitride (NbN),[7] and aluminum (Al),[8] as well as their granular and composite modifications, all of which offer large superconducting energy gaps and stable device performance.

Another key requirement for practical detector arrays is device miniaturization, which is largely determined by the footprint of individual resonators. Fractal resonator geometries offer a promising route toward increasing the packing density, as they significantly reduce parasitic coupling between adjacent segments compared to traditional meander structures. [9]

In our previous work, we investigated the response of a fractal MoRe superconducting resonator under continuous sinusoidal IR illumination.[10] In the present study, we examine the effect of pulsed IR radiation on the properties of a MoRe fractal resonator operating at 4.6 K.

## II. EXPERIMENT

A fractal microstrip resonator with lateral dimensions of approximately 3×3 mm$^2$ and a line width of 20 μm was fabricated from a 60 nm thick superconducting MoRe film deposited on a sapphire 0.5 mm substrate using standard photolithography. The topology of the resonator is shown in Fig. 1(a).

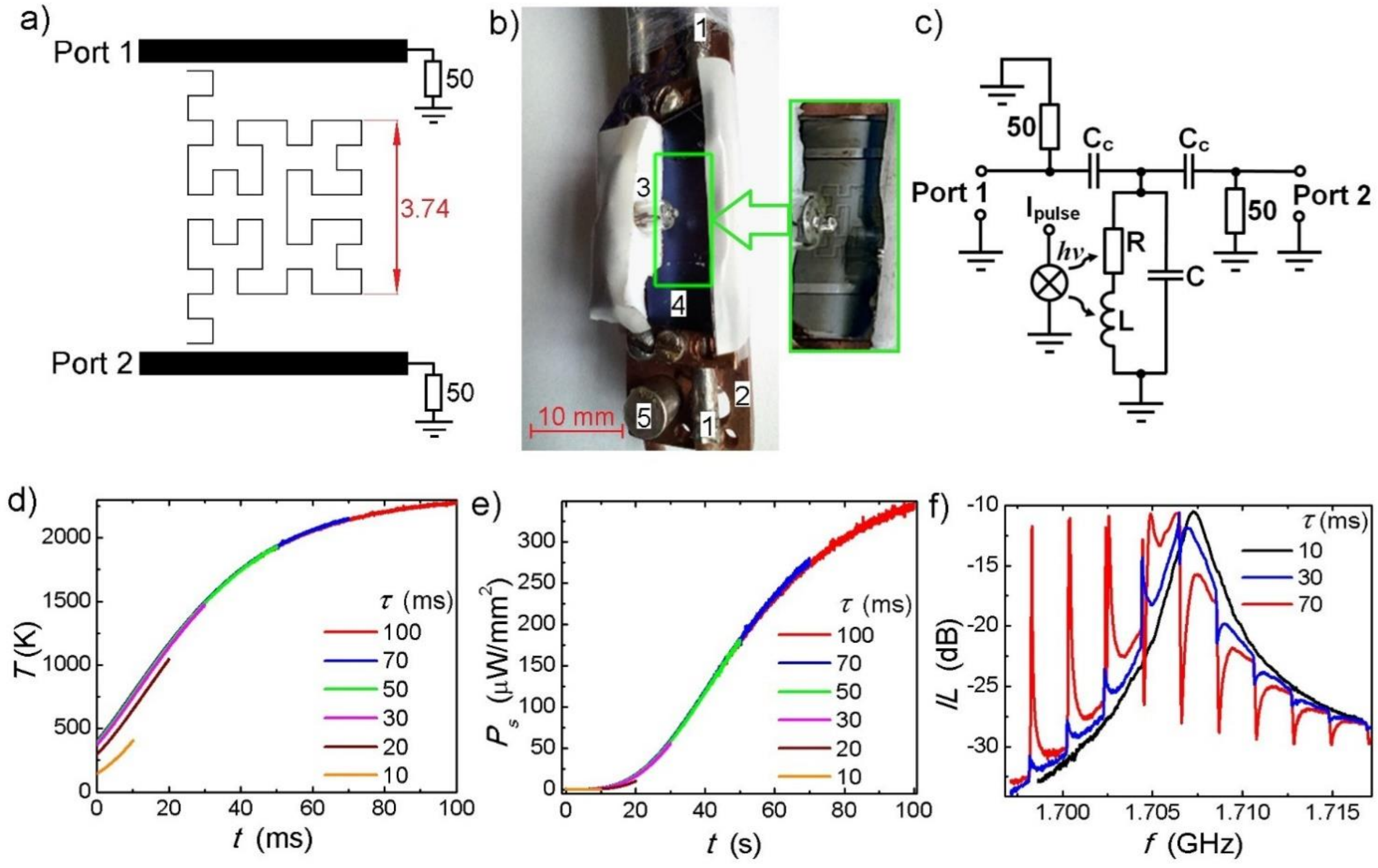

**FIG.1.** Topology of the fractal resonator with 50-Ω microstrip feed lines, capacitively coupled to the resonator, and configured for two-port (S-parameter) measurements (a). General view of the experimental setup (b). The system consists of: (1) a coaxial line, (2) a copper resonator holder, (3) an incandescent lamp used as the infrared source, (4) a silicon infrared filter, and (5) a thermometer. Inset (b) presents the MoRe fractal resonator without the silicon IR filter. Equivalent circuit of the measurement setup with a parallel *LCR* resonator capacitively coupled to the transmission lines. Pulsed infrared excitation ($h\nu$) induces changes in the resonator inductance $L$ and resistance $R$ (c). Temperature of the incandescent filament $T$ for pulses of different durations (d). IR power incident on the resonator $P_s$ (e). Series of amplitude–frequency response characteristics *IL* measured under IR pulses of varying duration (f).

The resonator was mounted on a copper sample holder and thermally anchored in helium vapor inside a 22-mm copper shield. As a source of infrared (IR) radiation, a miniature incandescent lamp was used and positioned 7.5 mm from the resonator (Fig. 1(b)). The incandescent lamp provided a broadband IR source that could be positioned inside the compact copper shield without the need for external optical coupling, enabling direct measurements in a liquid-helium transport dewar. A similar configuration was employed in Ref. 10. In the present configuration, a silicon wafer was placed between the lamp and the resonator to serve as an IR diffusing spectral filter. Since the glass envelope of the lamp blocks radiation at wavelengths longer than 4 µm,[11] and silicon is opaque for wavelengths shorter than 1 µm, the resonator was irradiated within the spectral range 1–4 µm. The final sample mounting configuration is shown in Fig. 1(b). In contrast to Ref. 10, the continuous sinusoidal excitation of the lamp was replaced with short current pulses supplied directly to the incandescent filament from a pulse generator. In Fig. 1(c), an equivalent circuit for two-port characterization of the resonator with capacitive coupling is presented. The resonator is represented by a parallel *LCR* model, where infrared pulse radiation originating from an incandescent lamp induces changes in the inductance $L$ and resistance $R$. Consequently, the resonance frequency $f_0$ shifts and the quality factor $Q_0$ of the resonator is modified under IR illumination. During each pulse, the voltage across the lamp and the current through it were recorded, which enabled determination of the time dependence of the filament resistance. Comparison of this resistance evolution with a previously obtained calibration curve of filament resistance versus temperature[6] yielded the filament temperature time profile $T(t)$. Figure 1(d) shows the resulting temperature dynamics for pulses of various durations, applied at a repetition rate of one pulse per second. Since the filament was operated in vacuum, its cooling occurred predominantly via thermal radiation. As a result, the filament did not fully cool down between pulses, and each subsequent pulse began at a temperature of approximately 400 K (see Fig. 1(d)). Thus, the preceding pulse established the initial conditions for the rapid heating of the filament during the next excitation. Figure 1(e) shows the time dependence of the radiation power incident on the microstrip film, taking into account the filter transmission, geometric factors, and the emissivity of the filament (details of the calculation are provided in Appendix A).

The resonator was implemented as a $\lambda/2$ open-ended fractal microstrip line segment (Fig. 1(a)) and was coupled to the external circuitry via symmetric capacitive coupling elements. The measurements were performed in a two-port configuration using a vector network analyzer. In this scheme, the complex transmission coefficient $S_{21}$ — defined as the ratio of the output signal amplitude at *port 2* to the input signal amplitude applied at *port 1* — was recorded as a function of frequency $f$ or time $t$.

In addition to the transmission coefficient, the insertion loss (*IL*) was evaluated, which characterizes the attenuation introduced by the resonator in the signal path. It is related to $S_{21}$ by $IL=-20\log_{10}|S_{21}|$. Thus, the measurement set yields both the complex transmission response $S_{21}(f,t)$ and the corresponding insertion loss $IL(f,t)$, providing information about the resonant frequency, quality factor, and temporal dynamics of the resonator.

Superconducting fractal microstrip resonators exhibit a pronounced nonlinear response to microwave excitation power.[12, 13] Therefore, all measurements in this work were carried out in the linear regime, using a low probe power of –36 dBm ($2.5\times10^{-7}$ W).

## III. RESULT AND DISCUSSION

Figure 1(f) shows a series of amplitude–frequency response curves of the resonator obtained under the influence of IR pulses repeated at a rate of one pulse per second during slow frequency scanning. The pulse duration was determined by the time during which current flowed through the incandescent filament of the lamp. As seen from the figure, pulses with duration of 10 ms have virtually no effect on the shape of the resonance curve, which is consistent with the absence of significant radiated power during the first 10 ms of the pulse (Fig. 1(e)). As the pulse duration increases, pronounced distortions of the resonance curve appear. During the pulse, sharp features are observed on the resonance profile, with their direction changing on the low- and high-frequency sides of the resonance. This behavior indicates a transient decrease in the resonance frequency at the moment of IR pulse excitation.

Next, the resonance frequency was fixed at $f=f_0$, and the parameter $S_{21}(t)$ was measured as a function of time $t$ under the influence of the IR pulses. Figure 2(a) shows a series of such time traces obtained for different pulse durations. A pronounced decrease in $S_{21}$ is observed during the interval in which current flows through the incandescent filament, followed by a slow relaxation. The latter is most likely associated with the continued IR irradiation while the filament cools down to a temperature of approximately 400 K (Fig. 1(e)).

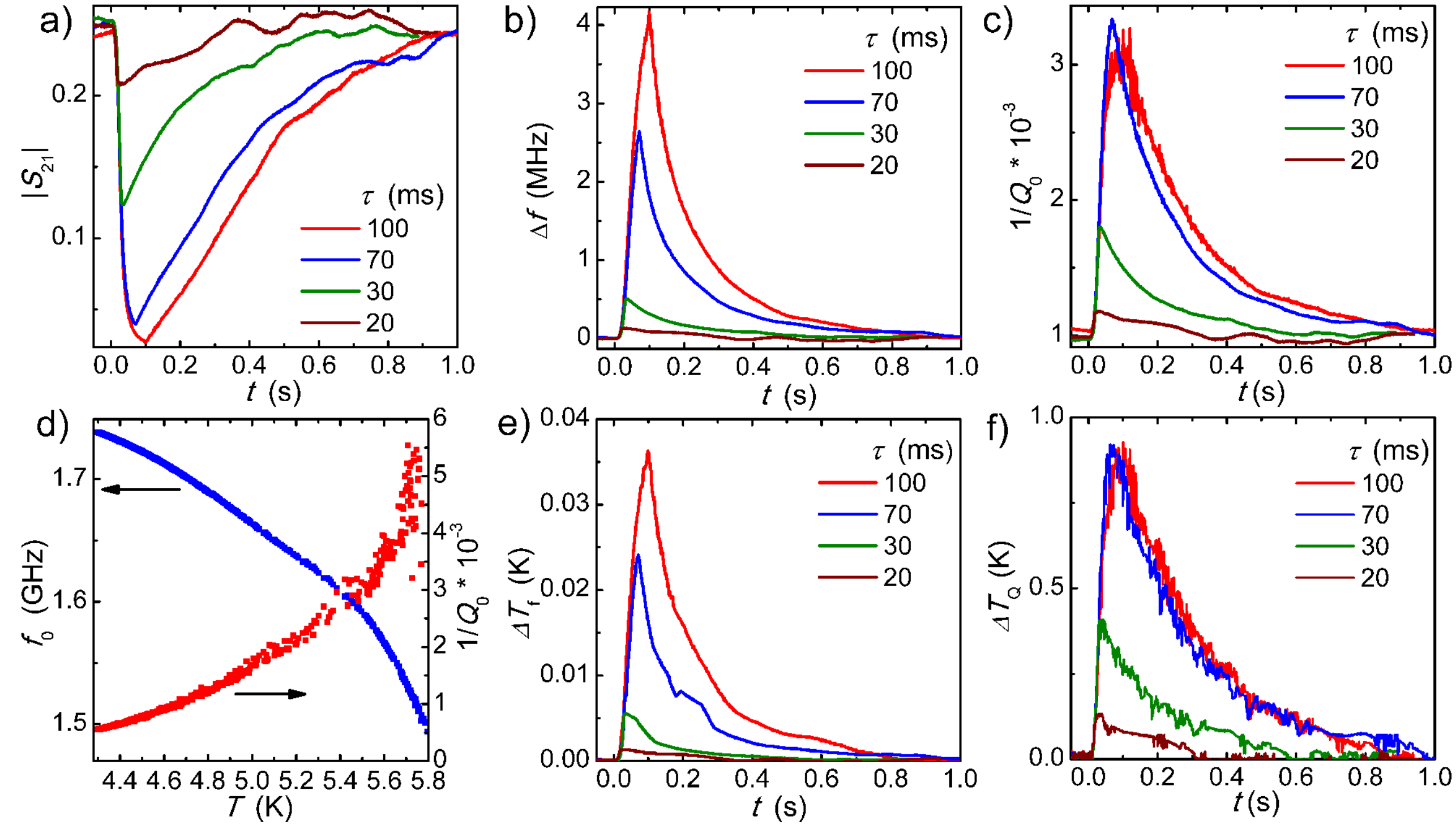

**FIG.2.** IR pulse amplitude response $|S_{21}|$ (a), resonance frequency shift $\Delta f$ (b), inverse quality factor $1/Q_0$ (c). Temperature dependences of the resonance frequency $f_0$ and inverse quality factor $1/Q_0$ (d),

and effective temperature rises derived from the resonance frequency shift $\Delta T_f$ (e) and from the variation of the quality factor $\Delta T_Q$ (f).

Using the procedure described in Appendix B, the measured time dependence $S_{21}(t)$ was converted into the temporal variations of the resonance frequency $\Delta f_0(t)$ and the inverse internal quality factor $1/Q_0(t)$ (Figs. 2(b) and 2(c)). In contrast to the resonance frequency shift, the inverse quality factor reaches saturation for pulse durations longer than approximately 70 ms.

Figure 2(d) shows the temperature dependences of the resonance frequency $f_0(T)$ and the inverse quality factor $1/Q_0(T)$, measured independently. By comparing these curves with the temporal dependences $\Delta f_0(t)$ and $1/Q_0(t)$, we determined the effective temperatures $T_f(t)$ and $T_Q(t)$ (Figs. 2(e) and 2(f)). These results show that for a 10-ms IR pulse, the maximal effective temperature changes are $\Delta T_f$=0.035 K and $\Delta T_Q$=0.87 K at $P_s$=340 μW/mm$^2$.

In Ref. 14, it was reported that when a film on a sapphire substrate (as in our case) is illuminated with optical radiation at a power density of $P_s$ = 1 mW/mm², its temperature increases by $\Delta T \approx 0.1$ K. This estimate is fully consistent with our experimentally inferred frequency-based temperature change of $\Delta T_f \approx 0.036$ K at $P_s$ = 340 μW/mm². At the same time, the dissipation-derived temperature shift $\Delta T_Q$ is much larger than $\Delta T_f$, demonstrating that the resonator response is dominated by non-equilibrium quasiparticle dynamics rather than by uniform thermal heating of the superconducting film.

Below we suggest an explanation of the observed dependences of the resonant frequency shift $\Delta f_0(P_s)$ and the resonator quality factor $\Delta(1/Q_0(P_s))$ on the incident radiation power as a result of the partial destruction of Cooper pairs and the creation of photoinduced quasiparticles, without taking into account the heating of the film.

Based on the temporal dependencies of $|S_{21}(t)|$, $\Delta f_0(t)$, and $1/Q_0(t)$ for $\tau$=100 ms (Figs. 2(a), 2(b) and 2(c)), as well as the absorbed power $P_s(t)$ (Fig. 1(e)), the following dependences were evaluated: dissipation responsivity $\Delta(1/Q_0(P_s))=1/Q_0(P_s)-1/Q_0(0)$; frequency responsivity $\Delta(f_0(P_s))=f_0(0)-f_0(P_s)$ ; transmission amplitude responsivity $\Delta|S_{21}(P_s)|=|S_{21}(0)|-|S_{21}(P_s)|$ (Fig. 3).

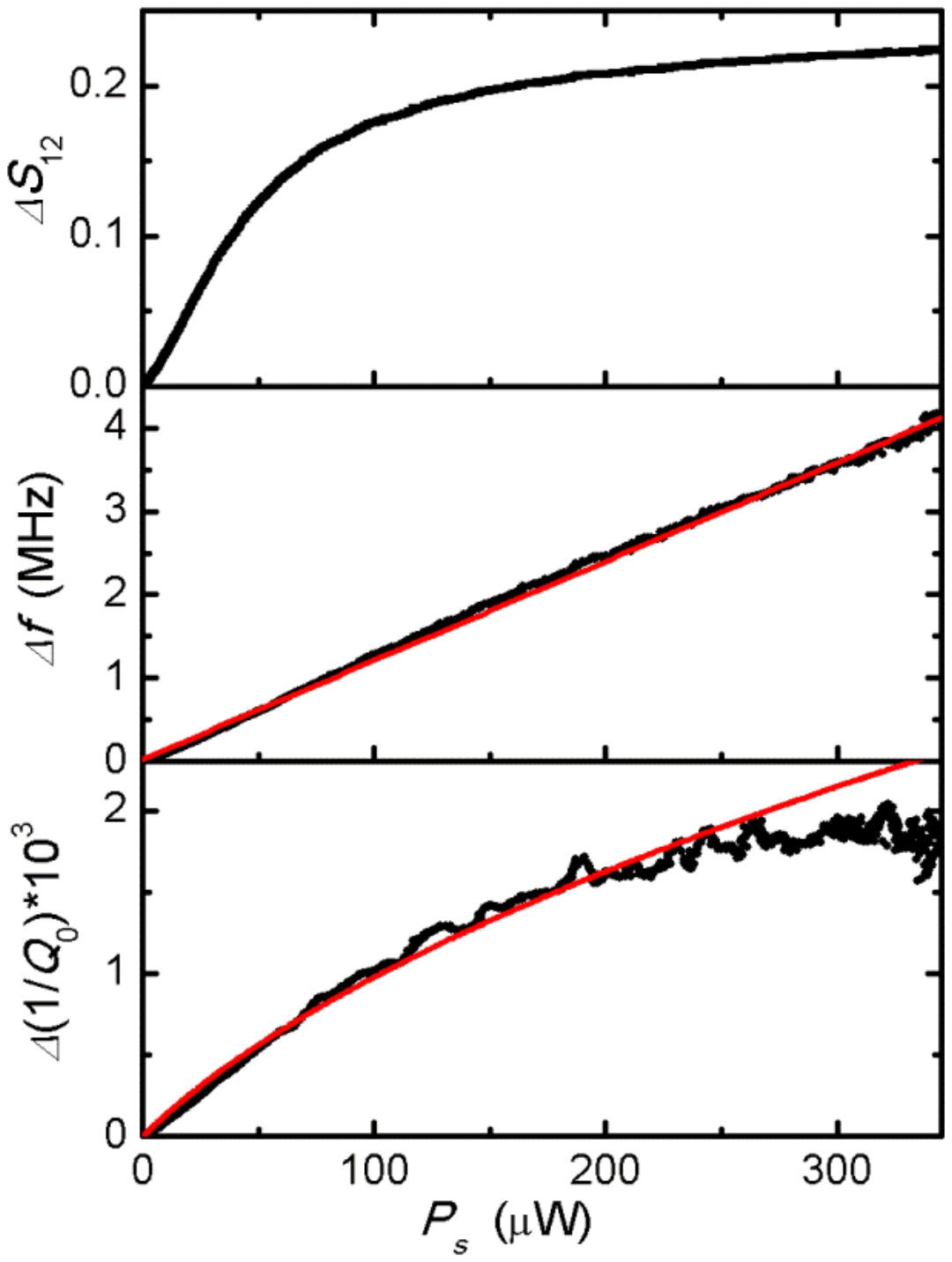

**FIG.3.** Dependence of the amplitude $\Delta S_{21}$, resonance frequency shift $\Delta f_0$ and dissipation $\Delta(1/Q_0)$ responsivities on the incident IR power (black symbols). The red solid line represents the fit obtained from theoretical dependencies given by Eqs. (4) and (7), respectively.

In what follows we analyze dependencies of the resonance frequency $f_0$ and the inverse quality factor $1/Q_0$ of the resonator under study on the incident radiation power $P_s$. We assume that the reason for these dependencies is related to the breaking of Cooper pairs in the superconductor by absorbed photons in the incident light with an energy above the threshold value $2\Delta$ ($\Delta$ is the gap in the quasiparticles energy spectrum in a superconductor).

The resonant frequency $f_0$ is determined by the well-known formula $f_0 = 1/(2\pi\sqrt{LC})$, where L and C are the characteristic inductance and capacitance of the resonator. For the superconducting thin film with a thickness $d \ll \lambda_L$ ($\lambda_L$ is the London penetration depth, $\lambda_L \sim 300$nm) one has for the kinetic inductance of the superconducting film $L \approx L_k = \mu_0(\lambda_L^2/d)$ (see, e.g. Ref. 15). This is just the case of our experiment. In the framework of the two-fluid model for superconductivity the London penetration depth is given by the relation: $\lambda_L^2(T) = m/\mu_0 e^2 n_s(T)$, where $n_s$ is the superfluid electron density, $e$ is the electron charge and $m$ is the electron mass.

Under the influence of light irradiation with a frequency above the threshold: $\hbar\omega > 2\Delta$, the destruction of Cooper pairs occurs and, accordingly, nonequilibrium quasiparticles (which are 'normal electrons' in the two-fluid model) are formed with a concentration $n_{ex}$, the value of which is proportional to the absorbed light power $P_s$, as follows from the simplest dynamic equation: $dn_{ex}/dt = I_0 - n_{ex}/\tau_r$, where $I_0$ is the rate of quasiparticle generation by absorbed light: $I_0 = \kappa P_s$ ($\kappa$ is a numerical coefficient), $\tau_r$ is the recombination time.

In the steady-state regime, $dn_{ex}/dt = 0$, and $n_{ex} = I_0\tau_r = \kappa P_s\tau_r \equiv \kappa_{ex} P_s$. Here $\kappa_{ex} = \kappa\tau_r$, - is the coefficient of proportionality between the concentration of nonequilibrium quasiparticles $n_{ex}$, and the absorbed radiation power $P_s$, in a steady state regime. The concentration of nonequilibrium quasiparticles is quite clearly proportional to the absorbed irradiation power $P_s$ (assuming that $\tau_r$ is independent on $P_s$). Further, we will assume that $n_{ex} \ll n_s(T)$ – equilibrium concentration of electrons in the superfluid condensate. According to the two-fluid model of a superconductor, such irradiation with a frequency above the threshold value should lead to a decrease in the concentration of superfluid electrons: $n_s \to n_s - n_{ex}$. Correspondingly, changes in the London depth, kinetic inductance, and resonant frequency occur:

$$\lambda_L^2 = \frac{m}{\mu_0 e^2 (n_s(T) - n_{ex})} = \lambda_L^2(T)\left[1 + \frac{n_{ex}}{n_s(T)}\right] = \lambda_L^2(T) + \Delta(\lambda_L^2)\,;\ \Delta(\lambda_L^2) = \lambda_L^2(T)\frac{n_{ex}}{n_s(T)} \qquad (1)$$

$$L_k = \mu_0 \left(\lambda_L^2/d\right) = L_k(T) + \Delta(L_k); \quad \Delta(L_k) = \mu_0 \left(\frac{\Delta(\lambda_L^2)}{d}\right) = L_k(T)\frac{n_{ex}}{n_s(T)}; \tag{2}$$

$$f = 1/\sqrt{LC} = 1/\sqrt{(L_k(T) + \Delta(L_k))C} = 1/\sqrt{L_k(T)C\left[1 + \frac{n_{ex}}{n_s(T)}\right]} \cong f_0(T)\left(1 - \frac{n_{ex}}{n_s(T)}\right); \tag{3}$$

$$\Delta f = f - f_0(T) = -f_0(T)\left(\frac{n_{ex}}{n_s(T)}\right) = -f_0(T)\left(\frac{\kappa P_s \tau_r}{n_s(T)}\right) = -f_0(T)\left(\frac{n_{ex}}{n_s(T)}\right) \propto P_s. \tag{4}$$

Thus, the shift in the resonant frequency is approximately proportional to the power of the incident light radiation in accordance with our experimental data (see Fig. 3). From Eq. (4) and experimental results presented in Fig. 3(b) it follows that in our experiments $\frac{n_{ex}}{n_s(T)} = \frac{\Delta f}{f_0(T)} \sim 10^{-3}$, in accordance with an assumption of a small amount of excited quasiparticles, stated above. The slope of $\Delta f(P_s)$ dependence in this figure determines the introduced above coefficient $\kappa_{ex}$, which relates the concentration of nonequilibrium quasiparticles, $n_{ex}$, and the absorbed radiation power, $P_s$, in a steady state regime: $\kappa_{ex} = \frac{1}{f_0}\frac{d\Delta f}{dP_S} n_s(T)d$. In our experiment $f_0 = 1.7\ GHz$; $\frac{d\Delta f}{dP_S} = 1.2 \cdot 10^4 \frac{Hz}{\mu W}$; $d = 60\ nm$. For chosen $n_s(T) = 10^{21}\ cm^{-3}$ one can estimate : $\kappa_{ex} = 4.2 \cdot 10^{10}\ cm^{-2}\mu W^{-1}$.

An explanation of the experimental dependence of the resonator inverse quality factor $1/Q_0$ on the light irradiation power $P_s$, shown in Fig. 3, requires a more in-depth analysis of the kinetics of irradiation-excited quasiparticles and their recombination. We will rely on the phenomenological model of Rothwarf and Taylor, [16] subsequently developed and supplemented in Refs. 17-18.

The value of $1/Q_0$ determines the value of the superconductor surface resistance in the microwave range: $1/Q_0 \propto R_s \propto \sigma_1$, where $\sigma_1$ is the real part of the high-frequency conductivity, proportional, according to the Drude formula, to the concentration of normal electrons. Therefore, a change in the concentration of normal electrons due to nonequilibrium quasiparticles induced by incident light leads to a change in the quality factor of the resonator: $\Delta(1/Q_0) \propto \Delta\sigma_1 \propto n_{ex}(P_s)$.

The phenomenological model approach, based on the work of Rothwarf and Taylor,[16] allows us to understand and explain the observed effect of saturation on the dependence of $\Delta(1/Q_0)$ on the incident light power $P_s$, as shown in Fig. 3.

The equations for the dynamics of nonequilibrium quasiparticles and phonons in a single-band superconductor exposed to light with a frequency ω > 2Δ/ħ, within the framework of the phenomenological model of Rothwarf and Taylor,[16] have the form:

$$\frac{dn_n}{dt} = I_0 - Rn_n^2 + \beta N_\omega\,; \tag{5}$$

$$\frac{dN_\omega}{dt} = R\frac{n_n^2}{2} - \beta\frac{N_\omega}{2} - \frac{N_\omega - N_{\omega,th}}{\tau_\gamma}, \tag{6}$$

where $n_n$ is the total concentration of quasiparticles: $n_n = n_n(T) + n_{ex}$, $I_0 = \kappa P_s$ is normalized incident power (see above), $R$ is the recombination coefficient, $N_\omega$ is the total number of phonons with energy $\hbar\omega > 2\Delta$, $N_{\omega,th}$ is the thermodynamically equilibrium number of phonons with the frequency $\omega$, $\tau_\gamma^{-1}$ is the net transition probability for phonons to be lost out of the energy range $\hbar\omega > 2\Delta$ by processes other than pair excitation, $\beta$ is the transition probability for the breaking of Cooper pairs by such phonons, and the $1/\gamma$ is the transition probability for phonons to escape from the energy range $\hbar\omega > 2\Delta$ by processes other than pair breaking. In the stationary case: $dn_n/dt = dN_\omega/dt = 0$, the solution of equations (5), (6) for the concentration of quasiparticles $n_n = n_n(T) + n_{ex}$ takes the form:[16]

$$n_n = \left[\frac{\beta N_{\omega,th} + I_0\left(1+\frac{\beta}{2}\tau_\gamma\right)}{R}\right]^{1/2}. \tag{7}$$

From (7) it follows that in the absence of IR irradiation ($I_0$=0) the equilibrium concentration of quasiparticles $n_n(T)$ equals $\left[\beta N_{\omega,th}/R\right]^{1/2}$. It also follows from expression (7) that for finite values of the incident irradiation power, the dependence of the concentration of photoinduced quasiparticles $n_{ex}(I_0)$ depends linearly on $I_0$ at low excitation levels, and at high values of $I_0$, this dependence changes to a flatter, square root dependence: $n_{ex}(I_0) \sim (I_0)^{1/2} = (\kappa P_s)^{1/2}$. This is in good agreement with the experimental results presented in Fig. 3, since, as noted above, $\Delta(1/Q_0) \propto \Delta\sigma_1 \propto n_{ex}(P_s)$. The solid line in Fig. 3 demonstrates the theoretical dependence which follows from the Eq.(7). One can see a rather good agreement of the theoretical curve with experimental data at moderate incident power levels. Meanwhile at higher power levels the there is an evident discrepancy between the theory, presented by Eq. (7), and experiment: the theoretical $\Delta(1/Q_0)$ dependence on $P_s$ is still growing up, meanwhile the corresponding experimental data on this type dependence demonstrate the tendency for the saturation. The possible saturation effect of the dependence $\Delta(1/Q_0)$ on the irradiation power at high values of $P_s$ can be due to the so-called effect of phonon bottleneck.[17-18] The latter corresponds to a regime in which the recombination of nonequilibrium quasiparticles is significantly slowed down due to the accumulation of high-energy phonons, which repeatedly break Cooper pairs.

As shown in Fig. 3, the frequency responsivity $\Delta f_0(P_s)$ increases approximately proportionally with the incident IR power $P_s$, indicating a monotonic increase in the kinetic inductance due to quasiparticle generation. In contrast, the dissipation responsivity $\Delta(1/Q_0(P_s))$ becomes nearly independent of power for $P_s$>250μW/mm$^2$. This saturation behavior correlates with the time-domain evolution of the inverse quality factor in Fig. 2(c), suggesting the establishment of a nonequilibrium steady-state quasiparticle population, in which additional absorbed power does not lead to further microwave losses. The combined evolution of $\Delta f_0(P_s)$ and $\Delta(1/Q_0(P_s))$ results in a pronounced $\Delta S_{21}$ amplitude responsivity at low radiation powers, followed by a gradual decrease for $P_s$>100μW/mm$^2$. Such behavior is consistent with a transition from a linear pair-breaking regime to a saturated dissipation regime, which may be associated with a quasiparticle relaxation bottleneck.

## IV. CONCLUSIONS

The resonator based on a MoRe superconducting film demonstrates a pronounced and nonlinear response to pulsed IR irradiation, governed primarily by nonequilibrium quasiparticle dynamics rather than by uniform thermal heating. Short IR pulses (10 ms) induce negligible changes in the resonance characteristics, consistent with the low radiated power during the initial heating stage of the filament. With increasing pulse duration and power, significant distortions of the resonance curve appear, reflecting a transient decrease in the resonance frequency due to enhanced kinetic inductance caused by quasiparticle generation.

Time-resolved measurements of $S_{21}(t)$, and their conversion into $\Delta f_0(t)$ and $1/Q_0(t)$, reveal fundamentally different behaviors for frequency shift and dissipation. While the resonance frequency shift scales approximately linearly with absorbed power, the inverse quality factor exhibits clear saturation for higher powers and longer pulse durations. This indicates the formation of a nonequilibrium steady-state quasiparticle population, in which further increases in absorbed power do not result in additional microwave losses.

The effective temperature analysis confirms that the dissipation-derived temperature rise significantly exceeds the frequency-derived temperature change, providing strong evidence that the detector response is dominated by quasiparticle pair-breaking processes rather than simple thermal effects. The observed saturation of dissipation responsivity and the reduction of transmission amplitude responsivity at higher powers suggest a transition from a linear pair-breaking regime to a saturated regime, likely associated with a quasiparticle relaxation bottleneck.

In the work of Goldie and Withington[19], it is shown that non-equilibrium phonons efficiently transfer their excess energy to the substrate on time scales much shorter than the experimental illumination time, and therefore do not lead to significant lattice heating. This conclusion is consistent with the small effective temperature changes of the superconducting film observed in our experiments.

Multiband superconductors, such as $MgB_2$[20], have attracted significant attention for MKID applications owing to the presence of multiple superconducting gaps, which can extend the dynamic range of detectable power and enhance resilience to background radiation. The coexistence of distinct superconducting gaps enables efficient quasiparticle generation over a broad range of photon energies and modifies non-equilibrium relaxation pathways, thereby improving detector performance under high optical loading.

Similar signatures of multigap superconductivity have also been reported in MoRe[21-23] alloys, indicating the coexistence of a small and a large superconducting gap. In this context, MoRe emerges as a highly attractive material platform for MKID development, as its intrinsic multiband character, combined with robust material properties, may enable enhanced power-handling capability and improved performance under strong infrared illumination.

## ACKNOWLEDGMENTS

This research was carried out within the framework of the project 2023.04/0157 funded by the National Research Foundation of Ukraine «High-speed matrix kinetic detector of long-wave infrared radiation».

## AUTHOR DECLARATIONS

### Conflict of Interest

The authors have no conflicts to disclose.

## DATA AVAILABILITY

The data that support the findings of this study are available within the article.

## APPENDIX A: RADIATION POWER

To determine the emitted radiation power, the temperature function $T(t)$ was substituted into Planck's law expression for the spectral radiance of an ideal blackbody. The power emitted per unit surface area in the wavelength interval $[\lambda_1, \lambda_2]$ was calculated as

$$P(\lambda_1, \lambda_2, T) = \int_{\lambda_1}^{\lambda_2} \frac{2\pi hc^2}{\lambda^5} \frac{1}{\exp\left(\frac{hc}{\lambda k_B T}\right)-1} d\lambda,$$

where $\lambda$ is the wavelength, $h$ is Planck's constant, $c$ is the speed of light, and $k_B$ is Boltzmann's constant. Using this expression, the time dependence of the emitted power in the spectral range 1–4 μm was obtained as a function $P(t)$.

To estimate the radiation power incident at the superconducting microstripe line, we write

$$P_s(t) = S_f \varepsilon \frac{1}{S_d} k_f P(t) = kP(t)\,,$$

where $S_f$=1.3×10$^{-6}$ m$^2$ is the area of the incandescent filament, $\varepsilon \approx 0.35$ is the emissivity of tungsten, $S_d$=7.07×10$^{-4}$ m$^2$ is the surface area of a sphere with radius equal to the distance between the lamp and the resonator $d$=7.5 mm, and $k_f$ is the transmission coefficient of the optical glass-silicon filter.

In the wavelength range 1–4 μm, both silicon and the glass envelope of the lamp are transparent; therefore, the dominant losses are associated with reflections at the vacuum–glass and vacuum–silicon interfaces. The glass envelope absorbs up to 10% of the emitted power.[11] Due to the high refractive index of silicon ($n$≈3.42), reflection at the two silicon–vacuum interfaces result in radiation losses of approximately 50%. Thus, the overall transmission coefficient is $k_f$≈0.45.

Substituting the values $S_f$, $S_d$, $\varepsilon$ and $k_f$ gives $k$=2.9×10$^{-10}$.

## APPENDIX B: RESONATOR TRANSMISSION MODEL

Near resonance, the transmission $S_{12}$ of a weakly coupled, symmetric two-port resonator can be written in the standard Lorentzian (input–output) form

$$S_{12}(\omega) = \frac{2\gamma_e}{\gamma_{tot} - 2j(\omega - \omega_0)}. \tag{B1}$$

where $\omega_0=2\pi f_0$ is the instantaneous resonant angular frequency,

$$\gamma_{tot} = \gamma_i + 2\gamma_e \tag{B2}$$

is the total decay rate - internal $\gamma_i$ plus two equal external channels decay rate $2\gamma_e$. The factor 2 in the numerator arises from the symmetric contribution of two couplers. The denominator $\gamma_{tot} - 2j(\omega - \omega_0)$ reflects damping and detuning of the resonant mode. At resonance $\omega = \omega_0$, in accordance with formula (B1) we obtain:

$$\gamma_e = \frac{S_{12}(\omega_0)}{2}\gamma_{tot}(\omega_0). \tag{B3}$$

The total decay rate $\gamma_{tot}$ is determined by the loaded quality factor $Q_{L0}$:

$$\gamma_{tot}(\omega_0) = \frac{\omega_0}{2Q_{L0}}. \tag{B4}$$

The external channels decay $\gamma_e$ can be found by formulas (B3) and (B4):

$$\gamma_e = \frac{S_{12}(\omega_0)}{2}\frac{\omega_0}{2Q_{L0}}. \tag{B5}$$

If the resonant frequency $\omega(t)$ and the total decay rate $\gamma_{tot}(t)$ vary over time $t$, while the measurements of $S_{21}(t)$ are performed at a fixed probing frequency $\omega_0$, then equation (B1) can be rewritten as:

$$\frac{1}{S_{21}(t)} = \frac{\gamma_{tot}(t)}{2\gamma_e} - j\frac{\omega_0 - \omega(t)}{\gamma_e}. \tag{B6}$$

By separating the real and imaginary parts of equation (B6), we obtain the expressions for determining the time-dependent functions $\gamma_{tot}(t)$ and $\Delta\omega(t)$:

$$\gamma_{tot}(t) = 2\gamma_e \Re[\frac{1}{S_{21}(t)}], \tag{B7}$$

$$\Delta\omega(t) = \omega_0 - \omega(t) = -\gamma_e \Im[\frac{1}{S_{21}(t)}]. \tag{B8}$$

In the approximation $\omega \approx \omega_0$, equation (B4) yields expressions for determining the loaded quality factor $Q_L(t)$ and the unloaded quality factor $Q_0(t)$:

$$Q_L(t) = \frac{\omega_0}{2\gamma_{tot}(t)}, \tag{B9}$$

$$\frac{1}{Q_0(t)} = \frac{1}{Q_L(t)} - \frac{4\gamma_e}{\omega_0} \tag{B10}$$

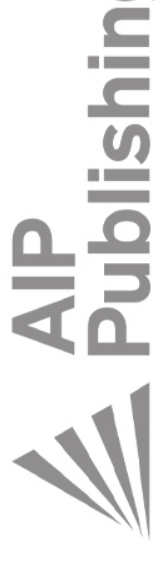